\documentstyle[preprint,aps]{revtex}
\widetext
\draft
\tighten

\begin{document}

\title{Action at a distance as a full-value solution of Maxwell equations:
basis and application of separated potential's method}

\bigskip

\author{{\bf Andrew E. Chubykalo and Roman Smirnov-Rueda}
\thanks{Instituto de Ciencia de Materiales, C.S.I.C., Madrid, Spain}}

\address {Escuela de F\'{\i}sica, Universidad Aut\'onoma de Zacatecas \\
Apartado Postal C-580\, Zacatecas 98068, ZAC., M\'exico}

\date{August 31, 1996}

\maketitle


\baselineskip 7mm

\begin{abstract}
The inadequacy of
Li\'{e}nard-Wiechert potentials is demonstrated as one of the examples
related to the inconsistency of the conventional classical
electrodynamics. The insufficiency of the Faraday-Maxwell concept to
describe the whole electromagnetic phenomena and the incompleteness of a
set of solutions of Maxwell equations are discussed and mathematically
proved. Reasons of the introduction of the so-called ``electrodynamics
dualism concept" (simultaneous coexistence of instantaneous Newton
long-range and Faraday-Maxwell short-range interactions) have been
displayed. It is strictly shown that the new concept presents itself as
the direct consequence of the complete set of Maxwell equations and makes
it possible to consider classical electrodynamics as a self-consistent and
complete theory, devoid of inward contradictions. In the framework of the
new approach, all main concepts of classical electrodynamics are
reconsidered.  In particular, a limited class of motion is revealed when
accelerated charges do not radiate electromagnetic field.

\end{abstract}


\newpage
\section{Introduction}

In the last century, the understanding of the nature of electromagnetic
phenomena was proceeding in a constant rivalry between two concepts of
interaction: namely, {\it Newton instantaneous long-range interaction}
(NILI) and {\it Faraday-Maxwell short-range interaction} (FMSI). At the
first moments, owing to the fundamental works of Gauss and Amp\'{e}re, all
electromagnetic phenomena were related to NILI. In other words, it was
understood that the interaction forces between both unmoving and moving
charges at some specific time were determined by their distribution and
the character of their motion at the same instant (implicit
time-dependence). As a matter of fact, the concept of field was merely
subsidiary (it was considered in a limited sense only as an external-force
field) and could be omitted at all.  On the contrary to this, the concept
of field is primary for FMSI, but charges and currents come to be
auxiliary. More fundamentally, a field is a system in its own right (has
physical reality), carries energy and fills the whole space. In accordance
with Faraday-Maxwell's idea, the interaction between charged particles can
be described only by the intermediary of a field as an energy-carrying
physical system. Any electromagnetic perturbation must be spread through
the space continuously from point to point during a certain amount of time
(finite spread velocity). Finally, the discovery of Faraday's law of
induction (explicit time-dependence of electromagnetic phenomena) and
the experimental observation of electromagnetic waves seemed to confirm
the field concept. Nevertheless, the idea of NILI still have many
supporters in our century. Among the physicists who have developed some
theories based, in any case, on this concept, we can find names such as
Tetrode and Fokker, Frenkel and Dirac, Wheeler and Feynman, Hoyle and
Narlikar [1]. This interest to the concept of NILI is explained by the
fact that classical theory of electromagnetism is an unsatisfactory theory
all by itself, and so there have been many attempts to modify either the
Maxwell equations or the principal ideas of electromagnetism. In
connection to this, we only mention some works which have tried to unify
 the advantage of the NILI concept with the conventional theory of field.
  They are the so-called ``retarded action-at-a-distance" theories [2-6].
   The fact that all new general solutions are represented by half the
    retarded plus half the advanced Li\'{e}nard-Wiechert solutions [7-8] of
     the Maxwell's equations makes it familiar with the conventional FMSI
      concept. On the other hand, these theories suggest the primacy of
       charge and use the notion of field as an external-force field like
        the action-at-a-distance theories. A single charged particle, in
         this approach, does not produce a field of its own, hence has no
           self-energy. Thus the classical theory can be saved from some
           difficulties like self-reaction force (self-interaction), the
            idea of a whole electromagnetic mass etc. It turns out,
            however, that no one effort to straighten out the classical
             difficulties has ever succeeded in making a self-consistent
              electromagnetic theory. Moreover, the principal difficulties
               in Maxwell's theory do not disappear still after the
                quantum mechanics modifications are made. In spite of the
                 great variety of methods applied to arrange the
                  situation, no one theory dealing with electromagnetism
                   had ever admitted the possibility of simultaneous and
                    independent co-existence of two types of interaction:
                     NILI and FMSI. A new approach, based on this idea,
                      has no need to modify neither Maxwell equations nor
                       the basis ideas of the classical electromagnetic
                        theory. In this work we take a complete set of
                         Maxwell equations as a correct one and show that
                          dualism of electromagnetic phenomena is an
                          intrinsic feature. Physical and mathematical
                          ground for that will be given in the next
                          sections.
\bigskip
\medskip

\section{Inadequacy  of  Li\'{e}nard-Wiechert  potentials.  A
paradox}

   The presence of a paradox in some theory no always means its
   inconsistency but often indicates the cause of difficulties. In this
   section we show one of the confusion's of classical electrodynamics in
   describing an electromagnetic field of an accelerated charge. The
   attractiveness of this example consists in the way it lightens the main
   conventional theory difficulties and the way it leads to the dualism
   idea.  Let us consider a charge $q$ moving in a laboratory  reference
   system with a constant acceleration $a$ along the positive direction of
   the $X$-axis. An electric field created by an arbitrarily moving charge
   is given by the following expression obtained directly from
   Li\'{e}nard-Wiechert potentials [9]:
   \begin{equation}
   {\bf E}(x,y,z,t)=q\frac{({\bf R}-R\frac{{\bf
         V}}{c})(1-\frac{V^{2}}{c^{2}})}{(R-{\bf R}\frac{{\bf
   V}}{c})^{3}}+q\frac{[{\bf R},[({\bf R}-R\frac{{\bf
   V}}{c}),\frac{{\bf{\dot{V}}}}{c^{2}}]]}{(R-{\bf R}\frac{{\bf
 V}}{c})^{3}}.
 \end{equation}
  We remind here that all values in the
   right-hand of (1) are taken in the moment of time $t_0=t-\tau$, where
   $\tau$ is the ``retarded time". We shall see that formula (1)
 satisfies
   the D'Alembert's equation along the
   $X$-axis at any time.  To begin with, we note that in a free space
   along the $X$-axis (except the site of a charge) an electric field
   component $E_x$ satisfies the homogeneous wave equation:
   \begin{equation}
  \Delta E_x
   -\frac{1}{c^2}\frac{{\partial}^{2} E_x}{{\partial t}^{2}}=0.
    \end{equation}
 To find the
  value $E_x$ at the moment of time $t$, one must take all the values
 in $rhs$ of (1) at the previous instant $t_0$ derived from the condition:
 \begin{equation}
 t_0=t-\tau=t-\frac{R(t_0)}{c};\quad\{(R^2=(x-x_0)^2+(y-y_0)^2+(z-z_0)^2)\}
 \end{equation}
(here $(x_0,y_0,z_0)$ is the site of the charge at instant $t_0$) or from
implicit function:
 \begin{equation}
F(x,y,z,t,t_0)=t-t_0-\frac{R}{c}=0.
\end{equation}
Then,  we have the following expression for $E_x(x,y,z,t)$:
\begin{equation}
E_x(x,y,z,t)=q\frac{\Bigl(x-x_0-R\frac{at_0}{c}\Bigr)\Bigl(1-\frac{a^2t_0^2}
{c^2}\Bigr)}{\Bigl(R-(x-x_0)\frac{at_0}{c}\Bigr)^3}-q\frac{a\Bigl((y-y_0)^2+
(z-z_0)^2\Bigr)}{c^2\Bigl(R-(x-x_0)\frac{at_0}{c}\Bigr)^3}.
  \end{equation}
   Substituting $E_x$ given by
 (5) in the wave equation (2), one ought to calculate in any case $\frac{
 \partial E_x}{\partial t_0}$,
  $\frac{\partial t_0}{\partial t}$ and $\frac{\partial t_0}{\partial x_i}$
 using differentiation rules for implicit function:  \begin{equation}
 \frac{\partial t_0}{\partial t}=-\frac{{\partial F}/{\partial t}}
 {{\partial F}/{\partial t_0}};\qquad \frac{\partial t_0}{\partial
 x_i}=-\frac{{\partial F}/{\partial x_i}} {{\partial F}/{\partial t_0}}.
 \end{equation}
  As a result of the substitution of (5) in (2) one obtains (one tends
 $y,y_0,z,z_0$ to zero after the differentiation):
   \begin{equation}
   \Delta E_x
   -\frac{1}{c^2}\frac{{\partial}^{2} E_x}{{\partial t}^{2}}=
0.
 \end{equation}
  This result is  not
reasonable if we remember that
wave equation (2) describes only transverse modes. In this particular
case, the $x$ component of electric field turns out to be the
longitudinal one and,  according to an ordinary wisdom,
is inconsistent with the wave equation (2).      Thus, the
   Li{\'{e}}nard-Wiechert potentials, as a solution of the complete set of
   Maxwell equations are inadequate for describing the properties of
   electromagnetic field along the direction of an arbitrarily moving
   charge. We note here that inadequacy of Li\'{e}nard-Wiechert
   potentials for describing the properties of relativistic fields was
   also shown by C.K. Whitney (see, e.g. [10]). The same singular behavior
   along $X$-axis direction displays another important quantity.  The
   Poynting vector represents the electromagnetic field energy flow per
   unit area per unit time across a given surface:  \begin{equation} {\bf
   S}=\frac{c}{4\pi}[{\bf E,H}];\qquad {\bf P}=\frac{1}{c^2}{\bf S},
   \end{equation} where {\bf S} is the Poynting vector, {\bf P} is the
   momentum density, {\bf E} and {\bf H}  are the electric and magnetic
   field strength, respectively.  One can easily see that expressions (8)
   are identically zero along the whole $X$-axis.  On the other hand, from
 the energy conservation law:  \begin{equation}
   w=\frac{E^2+H^2}{8\pi},\qquad \qquad \frac{\partial w}{\partial
   t}=-\nabla\cdot{\bf S} \end{equation} we conclude that $w$ and
 $\frac{\partial w}{\partial t}$  must differ from zero everywhere along
  $X$ and there is a linear connection between $w$ and $E^2$.  The
   conflict takes place, if, for instance, the charge is vibrating in some
   mechanical way along the $X$-axis, then the value of $w$ (which is a
   point function like $E$) on the same axis will be also oscillating.
   Then the question arises: {\it how does the point of observation, lying
   at some fixed distance from the charge on continuation of  X-axis,
   "know" about the charge vibration?}  The presence of  "retarded time"
   $\tau$ in (1) indicates that along the $X$-axis the longitudinal
   perturbation should be spread with the energy transfer (on the contrary
   to (8)). Since the vector {\bf S} is a product of the energy density
   and its spreading velocity {\bf v}:  \begin{equation} {\bf S}=w{\bf v}
\end{equation}
 then either the spreading velocity {\bf v} or the energy density $w$ must
   be zero along the $X$-axis. The first assumption puts aside the
   possibility of any interaction transfer. It is necessary to examine
   carefully the second one $(w=0)$. The Maxwell's equations state that
   time-varying fields are transverse. In electro- and magnetostatics (as
   correct stationary approximations of Maxwell's theory), the static
   fields are longitudinal, in the sense that the fields are derived from
   scalar potentials [11]. Consequently, we can assume the spreading of
   only longitudinal mods along the singular $X$-axis direction of
   our example, capable to change the field value in any point along this
   axis. In this case, according to (10), the energy of the longitudinal
   mods cannot be stored locally in space $(w=0)$ but the spread
   velocity may be whatever. On the other hand, FMSI concept forbids the
   spreading (not the presence) of any longitudinal electromagnetic field
   component in vacuum. Hence, this paradox can not be resolved in the
   framework of Faraday-Maxwell electrodynamics. This simplest example
   underlines the insufficiency of only transverse solutions of Maxwell's
   equations to describe full properties of electromagnetic field and
   leads directly to the dualism idea of {\it simultaneous and constant}
   co-existence of longitudinal ({\it action-at-a-distance}) and
   transverse electromagnetic interactions. In the next sections one can
   find mathematical and physical reasons for the dualism concept which
   permits to build up a self-consistent classical electrodynamics.  As a
   final remark, we make a reference to P.A.M. Dirac, who writes [12]:
   ``{\it As long as we are dealing only with transverse waves, we cannot
   bring in the Coulomb interactions between particles. To bring them in,
   we have to introduce longitudinal electromagnetic waves... The
   longitudinal waves can be eliminated by means of mathematical
   transformation. Now, when we do make this transformation which results
   in eliminating the longitudinal electromagnetic waves, we get a new
   term appearing in the Hamiltonian.  This new term is just the Coulomb
   energy of interaction between all the charged particles:  $$
   \sum_{(1,2)} \frac{e_1 e_2}{r_{12}} $$

... This term appears automatically when we make the transformation of the
elimination of the longitudinal waves}".

\bigskip
\medskip

\section{Reasons and foundations of the method of separated
potentials}

   Let us recall that a complete set of Maxwell equations is:
\begin{eqnarray}
&& \nabla\cdot{\bf E}=4\pi\varrho,\\
&& \nabla\cdot{\bf B}=0,\\
&& \nabla\times{\bf H}=\frac{4\pi}{c}{\bf j}+\frac{1}{c}\frac{\partial
{\bf E}}{\partial t},\\
&& \nabla\times{\bf E}=-\frac{1}{c}\frac{\partial {\bf B}}{\partial t}.
\end{eqnarray}
 If  this system of equations is really complete, it must describe all
   electromagnetic phenomena without exceptions.

       It is often convenient
   to introduce potentials, satisfying Lorentz condition:
\begin{equation}
\nabla\cdot{\bf A}+\frac{1}{c}\frac{\partial\varphi}{\partial t}=0.
\end{equation}
As  a result, the set of coupled first-order partial differential
equations (11)-(14) can be reduced to the equivalent pair of uncoupled
inhomogeneous D'Alembert's equations:
\begin{eqnarray}
&&\Delta\varphi-\frac{1}{c^2}\frac{{\partial}^{2}{\varphi}}{{\partial
t}^{2}}= -4\pi\varrho({\bf r},t),\\
&&\Delta{\bf A}-\frac{1}{c^2}\frac{{\partial}^{2}{\bf A}}{{\partial
t}^{2}}= -\frac{4\pi}{c}{\bf j}({\bf r},t).
\end{eqnarray}
 Differential
  equations have, generally speaking, an infinite number of solutions. An
   uniquely determined solution is selected by laying down sufficient
   additional conditions. Different forms of additional conditions are
   possible for the second order partial differential s equations: initial
   value and boundary conditions.  Usually, a general solution of
   D'Alembert's equation is considered as explicit time-dependent function
   $g({\bf r},t)$. In the stationary state the D'Alembert equation is
   transformed into the Poisson's equation which solution is everything an
   implicit-time dependent function $f({\bf R}(t))$.  Nevertheless, the
   conventional theory does not explain in details how the function
   $g({\bf r},t)$ is converted into implicit time-dependent function
   $f({\bf R}(t))$ (and vice versa) when the steady-state problems are
   studied.

       Further we shall demonstrate that
   former solutions of Maxwell's equations are incomplete and do not
   ensure a continuous transition between the D'Alembert and Poisson's
   equations solutions, respectively.  As a matter of fact, it will be
   shown that a mathematically complete solution of Maxwell's equations
   must be written as a linear combination of two non-reducible functions
   with implicit and explicit time-dependence:
\begin{equation}
f({\bf R}(t))+g({\bf r},t).
\end{equation}

     In the classical Faraday-Maxwell electrodynamics the Poisson's
   equation is mathematically exact for the steady-state problems. Based
   on the idea of a continuous nature of electromagnetic phenomena, one
   could suppose that the general solution of Poisson's equation should be
   continuously transformed to the D'Alembert's equation solution (and
   vice versa) when the explicit time-dependence appears (disappears).
   This requirement can also be formulated as a mathematical condition on
   the continuity of the general solutions of Maxwell's equations at every
   moment of time. By force of the uniqueness theorem for the second order
   partial differential equations, only one solution can exist satisfying
   the given initial and boundary conditions. Consequently, the continuous
   transition from the D'Alembert's equation solution into the Poisson's
   one (and vice versa) must be ensured by the continuous transition
   between the respective initial and boundary conditions. This is the
   point where FMSI concept fails. Really, only implicit time-dependence
   function $f({\bf R}(t))$ can be unique solution of Poisson's equation
   and boundary conditions for external problem are to be formulated in
   the infinity. On the other hand, the D'Alembert's equation solution is
   looking for only as explicit time-dependent function $g({\bf r},t)$
   since only that one corresponds to the classical FMSI concept as a
   physically reasonable solution. The boundary conditions in this case
   are given in a finite region. It has no sense to establish them at the
   infinity if it cannot be reached by any perturbation with finite spread
   velocity.  Dealing with large external region when the effect of the
   boundaries is still insignificant over a small interval of time, it is
   possible to consider the limiting problem with initial conditions for
   an infinite region (initial Cauchy's problem).

     Let us consider
   carefully the formulation of respective boundary-value problems in a
   region extending to infinity [13]. There are three external
        boundary-value problems for Poisson's equation. They are known as
        Dirichlet problem, Neumann problem and its combination. The
        mathematical problem, for instance, for the Dirichlet boundary
        conditions is formulated as follows. It is required to find the
        function $u({\bf r})$ satisfying:

\bigskip

          (i) Laplace's equation $\Delta u=0$  everywhere
        outside the given system of charges (currents);

\bigskip

        (ii) $u({\bf r})$ is
        continuous everywhere in the given region and takes the given
 value $G$ on the internal surface $S$: $u\vert_S  = G$;

\bigskip

             (iii) $u({\bf r})$
         converges uniformly to $0$ at infinity: $u({\bf r})\rightarrow 0$
         as $\vert{\bf r}\vert\rightarrow\infty$.

\bigskip

  {\it The final
condition} (iii) {\it is essential for a unique solution!}  In the case of
D'Alembert's equation the mathematical problem is formulated in a
different manner.  Obviously, we are interested only in the problem for an
infinite region (initial Cauchy's problem). So it is required to find the
function $u({\bf r},t)$ satisfying:
\bigskip

  (j) homogeneous D'Alembert's equation
        everywhere outside the given system of charges (currents) for
        every moment of time $t\geq 0$;
\bigskip

         (jj) initial conditions in all
        infinite region as follows:
 $$
 u({\bf r},t)\vert_{t=0}=G_1({\bf r});\qquad u_t({\bf r},t)\vert_{t=0} =
        G_2({\bf r}).
$$

          The condition (iii) about the uniform convergence at
        the infinity is not mentioned. We remind here that Cauchy's
   problem is considered when one of the boundaries is insignificant over
   all process time. This condition (iii) will never affect the problem
   and, hence, cannot be taken into account for the correct solution
   selecting.  However, it may be formally included into the mathematical
   formulation of D'Alembert's equation boundary-value problem to fulfil
   the formal continuity with Poisson's equation solution at the initial
   moment of time. Nevertheless, this condition is already meaningless the
   next instant of time since only explicit time-dependent solutions as
   $g({\bf r},t)$ (retarded solutions with finite spreading velocity) are
   considered.

     Thus, we underline here that the absence of the condition
   (iii) for every moment of time in the initial Cauchy's problem does not
   ensure the continuous transition into external boundary-value problem
   for Poisson's equation and, as a result, mutual continuity between the
   corresponding solutions cannot be expected by force of the uniqueness
   theorem. However, there is a way to solve the problem: to satisfy the
   continuous transition between the D'Alembert's and Poisson's equation
   solutions, one must look for a general solution in form of separated
   functions (18) non-reducible to each other. When applied to the
   potentials ${\bf A}$ and $\varphi$ this statement takes a form:
\begin{eqnarray}
&& {\bf A}={{\bf A}_0}({\bf R}(t))+{{\bf A}^{\ast}}({\bf r},t),\\
&& \varphi={{\varphi}_0}({\bf R}(t))+{{\varphi}^{\ast}}({\bf r},t).
\end{eqnarray}
In this case, the presence of the condition (iii) in the Cauchy's problem
   turns out to be meaningful for any instant of time, and the
   corresponding boundary conditions keeps continuity in respect of mutual
   transformation.

     As an additional remark, we conclude that the
   traditional solution of D'Alembert's equation cannot be complete, since
   the Faraday-Maxwell concept does not allow to take into account the
   first term in (18) as a valuable one at any moment of time. Turning to
   the previous section, we see that new solution in form (20) is able
   to change the electric field component $E_x$ along the $X$-axis at any
   distance and at any time. It is quite obvious now why Li'nard-Wiechert
   potentials (as only explicit time-dependent solution of Cauchy's
   problem) turned out not to be the complete solutions of Maxwell
   equations, and why they are not adequate to describe the whole
   electromagnetic field.

       Let us consider again the set of Maxwell's
   equations (11)-(14). A pair of uncoupled differential equations can be
   obtained immediately for the new general solution in form of
   separated potentials (19)-(20) (we omit boundary conditions
   premeditatedly):
\begin{eqnarray}
&& \Delta\varphi_0=-4\pi\varrho({\bf r},t),\\
&& \Delta{\bf A}_0=-\frac{4\pi}{c}{\bf j}({\bf r},t)
\end{eqnarray}
\qquad\qquad\qquad and
\begin{eqnarray}
&&\Delta\varphi^{\ast}-\frac{1}{c^2}\frac{{\partial}^{2}{{\varphi}^{\ast}}}
{{\partial t}^{2}}= 0,\\
&&\Delta{\bf A}^{\ast}-\frac{1}{c^2}\frac{{\partial}^{2}{{\bf
A}^{\ast}}}{{\partial t}^{2}}= 0.
\end{eqnarray}
 The  initial set of
Maxwell's equation has been decomposed into two independent sets of
   equations.  The first one (21)-(22) answers for the instantaneous
   aspect (``{\it action-at-a-distance"}) of electromagnetic nature while
   the second one (23)-(24) is responsible for explicit time-dependent
   phenomena.  The dualism as an intrinsic feature of Maxwell's
   equations is evident.  The potential separation (19)-(20) implies the
   same with respect to the field strengths:
\begin{eqnarray}
&& {\bf E}={\bf E}_0({\bf R}(t))+{\bf E}^{\ast}({\bf r},t),\\
&& {\bf B}={\bf B}_0({\bf R}(t))+{\bf B}^{\ast}({\bf r},t),
\end{eqnarray}
  where ${\bf E}_0$  and ${\bf B}_0$  are instantaneous (NILI) fields.  If
     we see again the formula (1)  based on Li\'{e}nard-Wiechert
     potentials, then in accordance with (25) the first term must be
     considered without ``retarded time" (at a given instant of time $t$)
        and the whole expression will be as follows:
\begin{equation}
{\bf E}[{\bf R}(t),{\bf R}_0(t_0),t_0]=q\frac{{\bf
R}(1-\frac{V^2}{c^2})}{R^2(1-\frac{V^2}{c^2}{\sin}^2\Theta)^{3/2}}+{\bf
E}^{\ast}[{\bf a}({\bf R}_0,t)],
 \end{equation}
  here $\Theta$ is the angle between the vectors {\bf V} and {\bf R}, ${\bf
 a}({\bf R}_0,t_0)$ is the acceleration of the charge $q$ in the previous
 moment of time $t_0=t-\tau$, $\tau$ is the ``retarded time". We note
 that the first term in (1) is mathematically equivalent to that in (27)
 (see [9]).  In the steady state ({\bf a}=0), the second term ${\bf
 E}^{\ast}$ must be zero, so (27) can be consistent with the requirements
 of the Lorentz transformation. The same approach is applicable to the
 Li\'{e}nard-Wiechert potentials. We leave out the complete modification
   of L.-W. potentials as well as an exact expression for ${\bf E}^{\ast}$
   which, while of interest in itselves, have not direct connection with
   the following material.

       To finish this section we conclude that NILI
   {\it must exist as a direct consequence of Maxwell equations}.
   According to this, both pictures, the NILI and the FMSI, have to be
   considered as two {\it supplementary descriptions of one and the same
   reality}. Each of the descriptions is only {\it partly} true. In other
   words, both Faraday and Newton in their external argument about the
   nature of interaction at a distance turned out right:  instantaneous
   long-rang interaction takes place, not {\it instead of}, but {\it along
   with} the short-range interaction in the classical field theory.

 \bigskip
 \medskip

\section{Relativistic  non-invariance of the concept  of energy of
    self-field of a charge (self-energy concept).  Mechanical analogy of
    Maxwell's equations}

    As a matter of fact, Maxwell's equations lend themselves to covariant
     description and are in agreement with the requirements of relativity.
      In the previous section we have not modified the set of Maxwell's
       equation, we have only separated two non-reducible
        parts in the general solution. Hence, the usual four-vector form
         of the basis equations can be used. For four-vectors of separated
         potentials we have automatically the following expressions:
\begin{equation}
\Box(A_{0\mu}+A^{\ast}_{\mu})=-\frac{4\pi}{c}j_{\mu};
 \qquad\qquad (\mu=0,1,2,3),
\end{equation}
where
\begin{equation}
A_{0\mu}+A^{\ast}_{\mu}=(\varphi_0+\varphi^{\ast},{\bf A}_0+{\bf
A}^{\ast});\qquad j_{\mu}=(c\varrho,{\bf j}).
\end{equation}

 To give some substance to the above formalism we exhibit explicitly the
 Poisson's equation for instantaneous four-vector ${\bf A}_{0\mu}$:
\begin{equation}
\Box A_{0\mu}=\Delta A_{0\mu}=-\frac{4\pi}{c}j_{\mu},
\end{equation}
where
\begin{equation}
A_{0\mu}=(\varphi_0({\bf R}(t)),{\bf A}_0({\bf R}(t))).
\end{equation}
  The Eq.(30) is an covariant also under Lorentz transformations. This is
   an exact consequence of (28) in the steady approximation. It is supported by the well-known fact that covariance
   is not necessary (it is sufficient) for the relativistic invariance.
   Nevertheless, in the Faraday-Maxwell electrodynamics this fact was
   always perceived as a quite odd one. Actually, potentials of an
   unmoving charge do not have explicit time-dependence. For a general
   Lorentz transformation from a reference system $K$ to an inertial system
   $K^{\prime}$ moving with the velocity ${\bf v}$ relative to $K$, the
   explicit time-dependence does not appear. Why those potentials keep
   implicit time-dependence under the Lorentz transformation? Without any
   approximation, the influence of a possible retarded effect is cancelled
   itself at any time and at any distance from the moving charge. From the
   other hand, the conventional theory is unable to describe correctly the
   transition from an uniform movement of a charge into an arbitrary one
   and then again into uniform over a limited interval of time. In this
   case, the first and the latter solutions at large distances can be
   given exactly by the Lorentz transformation. Furthermore the question
   arises:  what mechanism changes this potentials at the distance
   unreachable for retarded Li\'{e}nard-Wiechert fields? The lack of
   continuity between the corresponding solutions is obvious. It has the
   same nature as discussed in the above sections, due to incompleteness
   of existent solutions.

    The new approach also highlights the invariant deficiency of the
     self-energy concept in the framework of relativity theory. We confine
      our reasoning to the example of the electrostatic. The total
       potential energy of  $N$ charges due to all the forces acting
        between them is:
\begin{equation}
W=\frac{1}{2}\sum^{N}_{i=1}\sum_{j\neq i}\frac{q_i q_j}
{\vert r_i-r_j\vert}.
\end{equation}
 Here, the infinite self-energy terms $(i=j)$ are omitted in the double
  sum.  The expression obtained by Maxwell for the energy in an electric
  field, expressed as a volume integral over the field, is [14]:
\begin{equation}
W=\frac{1}{2}\int_{\cal V} E^2 d{\cal V}.
\end{equation}
 This corresponds to the Maxwell's idea that the system energy
   must be stored somewhere in space. The expression (33) includes
   self-energy terms and in the case of point charges they make infinite
   contributions to the integral. The introduction of a finite radius for
   the elementary charges enables to get rid of that difficulty but breaks
   down the possibility to see the classical electrodynamics as a
   self-consistent theory (Poincar\'{e}'s non-electrical forces [15]).

      In
   spite of introducing the self-energy concept long before the special
   relativity principle had arisen, there was no much alarm about the fact
    that it did not satisfy the relativity invariance condition. Strictly
      speaking, the Einstein's theory refutes the invariance of energy.
       The law of energy conservation cannot be maintained in its
        classical form. In a relativistically covariant formulation the
         conservation of energy and the conservation of momentum are not
          independent principles. In particular, the local form of
           energy-momentum conservation law can be written in a covariant
           form, using the energy-momentum tensor:
\begin{equation}
\frac{{\partial T}^{\mu\nu}}{{\partial x}^{\nu}}=0.
\end{equation}
   For an electromagnetic field, it is well-known that (34) can be
 strictly satisfied only for a free field (when a charge is not taken into
 account), whereas, for the total field of a charge this is not true,
 since (34) is not satisfied mathematically (four-dimensional analogy of
 Gauss's theorem). As everyone knows in classical electrodynamics, this
 fact gives rise to the ``electromagnetic mass" concept, which violates the
 exact relativistic mass-energy relationship $({\cal E}=mc^2)$. Let us
 examine this problem in a less formal manner. The equivalent
 three-dimensional form of (34) is the formula (9). The amount of
 electrostatic energy of an unmoving charge in a given volume ${\cal V}$
 is proportional to $E^2$ (see (33)).  According to (34) (or (9)), in a
 new inertial frame $K^{\prime}$, this value $W$ must be, generally
 speaking, an explicit time-dependent function ($\partial w/\partial t\neq
 0$).  Furthermore, this means also the explicit time-dependence for the
 electric field ($\partial {\bf E}/\partial t\neq 0$).  On the other hand,
 the electric field strength of an unmoving charge keeps its implicit
 time-dependent behavior under the Lorentz transformation ($\partial {\bf
 E}/\partial t=0$).  The conflict with the relativistic invariance
 condition is obvious.  The analogous reasoning can be applied for
 Coulomb's electrostatic energy of a system of charged particles. In this
 case, if one is thinking that electrostatic energy can be stored locally
 in space, the conflict with the relativity principle is inevitable.
 However, in the framework of the above-purposed separated potential's
 method it is possible to avoid those difficulties.  Actually, in the new
 general solution (25) it is ${\bf E}_0$ the only term exclusively linked
 to the charges.  According to the above speculation, no local energy
 conservation law can be written for this field ${\bf E}_0$. The
 mathematical form (32) must be saved for it. But there is no cause to
 reject the local form for the time-dependent free field ${\bf
 E}^{\ast}$.  In fact, the mathematical expression (33) is adequate for
 it. Thus, if one wishes not to get into trouble with the relativity
 principle, one must distinguish two different terms in the total electric
 field energy:
 \begin{equation}
W=\frac{1}{2}\sum^{N}_{i=1}\sum_{j\neq i}\frac{q_i q_j}
{\vert r_i-r_j\vert}+\frac{1}{2}\int_{\cal V} E^{{\ast}2} d{\cal V}.
 \end{equation}

We should make one further remark about this energy formula. In first
   place, the dualism concept reveals the dual nature of the
   electromagnetic field energy. So, for instance, the total electric
   energy is the electrostatic energy plus the electric energy of the free
   electromagnetic field. The first term is non-zero if the system
   consists of at least two interacting charged particles. The second term
   is taken as an integral over the region of ${\cal V}$ where the local
   value of $E^{\ast}$ is not equal to zero. In the next section the
   correctness of this energy representation for all electromagnetic field
   will be strictly verified by applying the principle of least action.
   The introduction of the self-energy concept in XIX century's physics
   can be explained historically. Maxwell considered the total
   electromagnetic field to be a uniform physical object on its own rights.

   Removing the self-energy concept, a valuable mechanical analogy of the
   Maxwell's equations in form of (21)-(24) can be used to understand why
   their general solution must be as separated  potential's
     (19)-(20). From the mathematical point of view, the two equations
      (21) and (22) correspond to the electro- and magnetostatic
       approximations, respectively, may be considered as wave equations
        with infinite spread velocity of longitudinal perturbations. If
         there is no local energy transfer, the Einstein's theory does not
          limit the signal spreading velocities. In this case, the set of
            differential equations for elastic waves in an isotropic media
            (see [16]) can be treated as mechanical analogy of (21)-(24):
\begin{eqnarray}
&& \frac{{\partial}^2 u_{\ell}}{{\partial t}^2}-
c^{2}_{\ell}\Delta u_{\ell}=0,\\
&& \frac{{\partial}^2 u_t}{{\partial t}^2}-c^{2}_{t}\Delta u_t=0.
\end{eqnarray}
  The general solution of (36)-(37) is the sum of two independent
   terms correspond to longitudinal $u_{\ell}$ and transverse $u_t$ waves:
 \begin{equation}
 u=u_{\ell}+u_t.
 \end{equation}
 If the longitudinal spreading velocity approaches formally to infinity
   ($c_{\ell}\rightarrow\infty$) then (36) transforms into Laplace's
   equation whereas the function $u_{\ell}$ turns out to have an implicit
   time-dependence.  Thus, the formula (38) takes the form of separated
   potential's solution (19)-(20).

       To end this section, we note that the
   idea of non-local interactions can be immediately derived from the
   Maxwell's equations as an exact mathematical result. On the other hand,
   some of the quantum mechanical effects like Aharonov-Bohm effect [17],
   violation of the Bell's inequalities [18-19] etc. point out indirectly
   to the possibility of non-local interactions in electromagnetism.
   Nevertheless, in this work we prefer to confine themselves to the
   classical theory.
\bigskip
\medskip

\section{Hamiltonian  form  of  Maxwell's  equations  from  the
point of view of  separated  potential's  method}

   In the latter section we have introduced the prototype for a new
   electromagnetic energy interpretation. In this section we shall discuss
   general field equations for arbitrary fields from the standpoint of the
   principle of least action and the change in their interpretation due to
   the new dualism concept. In
   extending the separated potential's method no modifications at all are
   necessary in the set of Maxwell's equations to make them agree with the
   requirements of the covariant formulation. Hence, in the steady
   approximation ($\varphi^{\ast}=0,  {\bf A}^{\ast}=0$) a relativistic
   action for a system of interacting charged particles can be written in
   conventional form [9]:
\begin{equation}
S_{m}+S_{mf}=\int\Bigl(-\sum^{N}_{a=1}m_acds_a-
\sum^{N}_{a=1}\frac{e_a}{c}\sum_{\mu=0}A_{0(\mu a)}dx^{\mu}_{a}\Bigr),
\end{equation}
where $A_{0(ma)}$ is the {\it instantaneous} potential ($\varphi_0,
{\bf A}_0$) in the four-point on the world-line of the particle with the
number ``$a$" created by other particles. This expression is sufficient to
derive the first couple of equations (21)-(22) from the least action
principle.  It can be proved directly rewriting the second term in (39)
as:  \begin{equation}
S_{mf}=-\frac{1}{c}\int\sum_{\mu}A_{0\mu}j^{\mu}d{\cal V}dt,
\end{equation}
using the Dirac's expression for four-current:
\begin{equation}
j_{\mu}({\bf r},t)=\sum_{a}\Bigl[\frac{e_a}{4\pi}\Delta
\Bigl(\frac{1}{\vert {\bf r}-{\bf r}_a\vert}\Bigr)\Bigr]u_{\mu a},
\end{equation}
where $u_{\mu a}$  - four-velocity of the charged particle ``$a$".
   Generally, for a system of arbitrary moving charges, the time-dependent
   potentials ($\varphi^{\ast},{\bf A}^{\ast}$) appear in the general
   solution.  It means that an additional term corresponding to the free
   electromagnetic field must be added to (39). In the first place, it
   must vanish under the transition to the steady approximation
    ($\varphi^{\ast}=0,{\bf A}^{\ast}=0$).
    On the other hand, the variation of this term has to lead to the
   second pair of equations (23)-(24).  As a result, it is easy to see
   that conventional Hamiltonian form can be adopted to describe the
           presence of the free electromagnetic field [9]:
\begin{equation}
S_f=-\frac{1}{16\pi}\int\sum_{\mu,\nu}F_{\mu\nu}F^{\mu\nu}d{\cal V}dt,
\end{equation}
where
\begin{equation}
F_{\mu\nu}=\frac{\partial A^{\ast}_{\nu}}{\partial x^{\mu}}-
\frac{\partial A^{\ast}_{\mu}}{\partial x^{\nu}}.
\end{equation}
Finally, it remains to be proved that from the variation derivative:
\begin{equation}
\delta S_f=-\int\sum_{\mu}\Bigl(\frac{1}{4\pi}\sum_{\nu}\frac
{\partial F^{\mu\nu}}{\partial x^{\nu}}\Bigr)\delta A^{\ast}_{\mu}
d{\cal V}dt
\end{equation}
one obtains the covariant analogous of (23)-(24) in
the following form:
\begin{equation}
\sum_{\nu}\frac{\partial}{\partial x^{\nu}}F^{\mu\nu}=
\sum_{\nu}\frac{\partial}{\partial x^{\nu}}
\Bigl[\frac{\partial A^{\ast\nu}}{\partial x_{\mu}}-
      \frac{\partial A^{\ast\mu}}{\partial x_{\nu}}\Bigr]=0.
\end{equation}
The only difference with the classical field interpretation consists in
   the way how electromagnetic potentials take part in this Hamiltonian
   formulation. Actually, the second term in (39) contains only
   instantaneous potentials whereas $S_f$ is related with time-varying
   field components. Consequently, contrary to traditional interpretation,
   the quantity $F^{\mu\nu}$ can be defined as a free electromagnetic
   field tensor.

   In the light of the new approach, the electromagnetic
   energy-momentum tensor demands some corrections in the interpretation
   of its formal mathematical formulation [9]:
\begin{equation}
T^{\mu\nu}=-\frac{1}{4\pi}\sum_{\rho}F^{\mu}_{\rho}F^{\nu\rho}+
\frac{1}{16\pi}{\sl g}^{\mu\nu}
\sum_{\beta,\gamma}F_{\beta\gamma}F^{\beta\gamma}.
\end{equation}
As a consequence of (43), in this form it can describe the energy-momentum
conservation law for, exclusively, free electromagnetic field as follows:
\begin{equation}
\sum_{\nu}\frac{\partial T^{\mu\nu}}{\partial x^{\nu}}=0
\end{equation}
which supports the new interpretation of electric field energy given in
the previous section.  Strictly speaking, from the point of view of the
dualism concept, the total field energy $W$ must consist of two
non-compatible parts: on one hand, the energy $W_{mf}$  of electro- and
magnetostatic interaction between charges and currents ({\it non-local}
term), on the other hand, the energy $W_f$  of the free electromagnetic
field ({\it local} term):
\begin{equation}
W=W_{mf}+W_f.
\end{equation}
It contradicts considerably the FMSI concept about the unique nature of
   electromagnetic field energy.  Summarizing these results we see that
   the concept of potential ({\it non-local}) energy and potential forces
   must be conserved in the classical electrodynamics as valid. So, the
   system of charges and currents in the absence of free electromagnetic
   field must be considered as conservative system without any
   idealization.  As an important remark we note the physical meaning of
   Poynting vector has been changed notably. So far the classical theory
   dealt with it as a quantity attributed to all dynamic properties of the
   total electromagnetic field. From the new point of view, it can be
   non-zero  only in the presence of {\it free} field. The great problem
   of the classical electrodynamics, the indefiniteness in the location of
   the field energy, does not exist anymore. In particular, the flux of
   the electromagnetic energy in the steady state has no sense since no
   presence of the free electromagnetic field is supposed in this
case.
\bigskip
\medskip

\section{Non-radiation condition for  free electromagnetic
field}

   In this section we shall discuss the energy balance between the system
   of interacting charged particles and free electromagnetic field, namely,
   energy and momentum lost by radiation. Turning to the latter section
   results we must examine carefully one essential difference in
   electromagnetic energy interpretation. Let us write the total
   relativistic action as:
\begin{equation}
S=S_m+S_{mf}+S_f.
\end{equation}
Although we adopt the same denominations used in the conventional theory,
   the physical essence of the last two terms has changed significantly.
   Usually, the interaction between particles and electromagnetic field
   was attributed to $S_{mf}$ whereas the properties of electromagnetic
   field manifested itselves by the additional term $S_f$.

       In the new
   approach, no concept of field as intermediary is needed to describe the
   interaction between charges (currents). Hence, $S_{mf}$ cannot be
   treated in terms of particle-field interaction. Such interaction as
   well as the intrinsic properties of a free electromagnetic field are
   enclosed now in the last term $S_f$. The possible free field
   interaction with the system of charges (currents) depends entirely on
   its location in space.  This reasoning makes it possible to consider
   the isolated system of charged particles and free field  as consisting
   of two corresponding subsystems. Each of the subsystems may be
   completely independent if there is no mutual interaction  (for
   instance, free electromagnetic field is located far from the given
   region of charges and currents). In the steady approximation  the
   first subsystem (charges and currents) can be considered as
   conservative. In other words, it means the total Hamiltonian of the
   whole isolated system can be decomposed into two corresponding parts:
\begin{equation}
{\cal H}={\cal H}_1+{\cal H}_2,
\end{equation}
where  ${\cal H}_1$  is the Hamiltonian of the conservative system of
  charges and currents. It involves apart from electro- and magnetostatic
  energy also mechanical energy of particles (corresponds to the action
  $S_m + S_{mf}$).  ${\cal H}_2$  is the Hamiltonian of the free
  electromagnetic field (corresponds to the action $S_f$).

    We remind here
  that in the relativistic case, the energy is the zero component of the
  momentum.  However, if we deal with the isolated system, the total
  Hamiltonian is not time-dependent and the energy conservation law as
  well as the momentum conservation may be treated independently. It is
  important to note that such separation into two subsystems is valid only
  in the new approach.  The conventional interpretation of $S_f$ did not
  allow to consider it apart. Actually, in the steady approximation $S_f$
  was not zero, and corresponded to the self-energy of field [9]:
\begin{equation}
S_f=\int^{t_1}_{t_2}{\cal L}_{f}dt,
\end{equation}
where
\begin{equation}
{\cal L}_f=\frac{1}{8\pi}\int_{\cal V}(E^2-B^2)d{\cal V}.
\end{equation}
Here  ${\bf E}$ and ${\bf B}$ are the total electro- and magnetic field
   strengths, respectively. Thus, the fact that $S_f$  is responsible
   solely for free field, turns out to be a meaningful argument in
   separating into two subsystems.  It is often possible to extract a
   large amount of information about the physical nature of the system
   using conservation laws, even when complete solutions cannot be
   obtained. Let us now consider the case when charges (currents) and free
   electromagnetic field are located in the same region and become
   interacting.  Internal forces of mutual reaction between two subsystems
   are usually named as internal dissipative forces. They carry out the
   energy exchange inside the total isolated system. In terms of the
   Hamiltonian formalism it can be expressed as a corresponding
   Hamiltonian evolution  (see, for instance, [20]):  \begin{equation}
\frac{d{\cal H}_{1,2}}{dt}=\frac{\partial {\cal H}_{1,2}}{\partial t}+
{\cal P}^{\tt ext}_{1,2}+{\cal P}^{\tt int}_{1,2},
\end{equation}
where ${\cal P}_{1,2}^{\tt ext}({\cal P}_{1,2}^{\tt int})$ is the power of
the external (internal) forces acting on two the subsystems, respectively.
In our case ${\cal P}_{1}^{\tt ext}$ and ${\cal P}_{2}^{\tt ext}$ appear as
 a result of the mutual interaction.  On the other hand, any internal
non-potential force in the first subsystem can also cause energy
dissipation (${\cal P}_{1}^{\tt int}$).  Even in the absence of a real
mechanical friction, other internal non-potential forces (for example,
inhomogeneous gyroscopic forces) can still act in this subsystem and
dissipate energy.  In other words, if initially there is no free
electromagnetic field (${\cal H}_2=0$), it can be created by internal
non-potential forces (${\cal P}_{1}^{\tt int}$) acting in the first
subsystem (${\cal H}_2$ stops to be zero).  It means that energy is lost
by radiation in the subsystem of charges and currents.  In mathematical
language the corresponding energy balance can be written as follows:
\begin{equation}
\frac{d}{dt}({\cal H}_1+{\cal H}_2)=
\dot{\cal H}_1+\dot{\cal H}_2=0,
\end{equation}
     where $\dot{\cal H}_1$ and $\dot{\cal H}_2$  are energy change rates
   for the first and the second subsystems respectively. It might be
   easily noted that the energy balance (54) is symmetrical in respect to
   time reversion which is in accordance with the time symmetry of
   Maxwell's equations.  The real direction of the energy exchange process
   may be determined by some subsidiary conditions. On the contrary to
   this, the energy balance in the conventional electrodynamics was always
   irreversible in time. From the other hand, the former class of theories
   based on the action-at-a-distance principle (for example, the
   electrodynamics of Wheeler and Feynman) did not consider at all the
   third term $S_f$  in (49), corresponding to radiation reaction. As a
   matter of fact, there were no radiation effects in those theories, but
   only interactions of a number of particles.

     To end the section we
   formulate the previous statement about the energy conservation as
 {\bf the condition of non-radiation of the free electromagnetic field}:

     {\it If
   in an isolated system of charges (currents) in the absence of free
   electromagnetic field (${\cal H}_2=0$), all internal non-potential
   forces are compensated or do not exist then this system will not
   produce (radiate) free electromagnetic field (${\cal H}_2$ remains
   zero) and will keep  conservative system itself}.

       This implies not only an equilibrium
   between radiation and absorption but no radiation at all. As a simple
   example of a non-radiating system we can consider two charged particles
   moving with acceleration along a direct line under mutual Coulomb
   interaction.  The absence of other frictional forces is supposed. The
   presence of any inhomogeneous gyroscopic (Lorentz-type) forces here are
   not expected due to the one-dimensional character of motion.  Some
   mention should also be made of the many-particle system. It is possible
   that there is some limited class of motion when all non-potential (for,
   example, internal gyroscopic forces) can be compensated due to the own
    magnetic moment of charged particles. This possibility would be of
    particular interest in the attempt to understand the quantum mechanics
    principles.

   In the present work the question of interaction of free fields with
   sources (charges and currents) is given in perfunctory manner and
   should be studied carefully. It should be  compared with
   the older non-radiation theories based on the {\it extended} Dirac
   electron models (see, for instance, [21]).  Furthermore, emission,
   absorption and, for instance, scattering processes can be caused by
   the interaction of matter fields with the $B(3)$ spin field. It is
   created by transverse left- and right- circular polarized waves, as
   found by Evans and Vigier [22-25].  On the other hand, the existence of
   the longitudinal  $B(3)$ field may hint on non-zero photon mass.
   Theoretical constructs of such a type were introduced and developed by
   Einstein, Schr\"odinger, Deser, de Broglie and Vigier (see, e.g.
   [26]).  However,  relations between $B(3)$ and other longitudinal
   solutions of Maxwell equations, as well as the problem of photon mass,
   must be studied more carefully.

\bigskip
\medskip

\section{Conclusions}

    Finally, we conclude that the FMSI concept could not give a complete
   and adequate description of the great variety of electromagnetic
   phenomena. It has been shown that other concept (the so-called
   {\it dualism concept}), consistent with the full set of Maxwell's
   equations, can be accepted as a correct description of
   electromagnetism. In other words, the new concept states that there is
   a {\it simultaneous} and {\it independent} coexistence of Newton
   instantaneous long-range (NILI) and Faraday-Maxwell short-range
   interactions (FMSI) which cannot be reduced to each other. The reasons
   are based on the mathematical method (so-called {\it separated
   potential's method}) proposed in this work for a complete general
   solution of Maxwell's equations. As a result, the incompleteness of
   former solutions of Maxwell's equations is proved.

     In the framework
   of the new approach, all main concepts of the classical electrodynamics
   have been reconsidered.  In particular, it has been shown that the dual
  nature of the total electromagnetic field must be taken into
  consideration. On one hand, there is a free electromagnetic field
  ${\bf E}^{\ast} ({\bf B}^{\ast})$ which has no direct connection with
  charges and currents, and can be transferred {\it locally}. On the other
  hand, there is a field ${\bf E}_0 ({\bf B}_0)$ linked exclusively to
  charges (currents) and responsible for interparticle interaction which
  {\it cannot be transferred locally} in space.  However, in total, this two
  kinds of electromagnetic field ${\bf E}_0+{\bf E}^{\ast}$ $({\bf
  B}_0+{\bf B}^{\ast})$ as a superposition satisfy Maxwell's equations and
  are observed experimentally as an unique electromagnetic field.  Other
  quantities of the classical electrodynamics such as electromagnetic
  field tensor, electromagnetic energy-momentum tensor etc. have also
  changed their physical meanings.  In particular, the Poynting vector can
  be associated {\it only with the free electromagnetic field}. In the
  light of this result, the problem of the indefiniteness in the field
  energy location has no place and no flux of electromagnetic energy in
  steady state can be derived from the theory.  Also problems such as
  {\it self-force, infinite contribution of self-energy, the concept of
   electromagnetic mass, radiation irreversibility in time with respect
   of time symmetry of Maxwell equations} have been removed in the new
   approach.  A new interpretation of the energy conservation law is
   possible as a non-radiation condition which states that {\it a limited
   class of motion exists when accelerated charged particles do not
   produce electromagnetic radiation}.

\bigskip
\medskip

{\large Acknowledgments}

\bigskip

We are grateful to Dr. V. Dvoeglazov and Professor J. V. Narlikar for many
stimulating discussions.  We acknowledge papers of Professor M. W. Evans,
which gave support to our belief that ideas presented here have sufficient
grounds.  Authors are indebted for financial support, R. S.-R., to the
Comunidad de Madrid, Spain, for the award of a Postgraduate Grant, A. Ch.,
to the Zacatecas University, M\'exico, for a Full Professor position.

\end{document}